\documentclass[twocolumn,amsmath,aps]{revtex4}
\usepackage{graphicx,color}
\usepackage{CJK}
\usepackage{bm}
\usepackage[hypertex]{hyperref}

\newcommand{\be}{\begin{equation}}
\newcommand{\ee}{\end{equation}}
\newcommand{\bea}{\begin{eqnarray}}
\newcommand{\eea}{\end{eqnarray}}
\newcommand{\bsube}{\begin{subequations}}
\newcommand{\esube}{\end{subequations}}

\newcommand{\Eq}[1]{Eq.\,(\ref{#1})}

\newcommand{\dg}{\dagger}
\newcommand{\la}{\langle}
\newcommand{\ra}{\rangle}

\newcommand{\nl}{\nonumber \\}

\newcommand{\beq}{\begin{equation}}
\newcommand{\eeq}{\end{equation}}
\newcommand{\beqn}{\begin{eqnarray}}
\newcommand{\eeqn}{\end{eqnarray}}
\newcommand{\bsub}{\begin{subequations}}
\newcommand{\esub}{\end{subequations}}

\begin{document}
\begin{CJK*}{GBK}{Song}

\title{Transport probe of nonadiabatic transition caused by Majorana moving}

\author{Luting Xu}
\email{xuluting@tju.edu.cn}
\affiliation{Center for Joint Quantum Studies and Department of Physics,
School of Science, Tianjin University, Tianjin 300072, China}

\author{Xin-Qi Li}
\email{xinqi.li@tju.edu.cn}
\affiliation{Center for Joint Quantum Studies and Department of Physics,
School of Science, Tianjin University, Tianjin 300072, China}

\date{\today}

\begin{abstract}
We propose a transport probe scheme to detect the
nonadiabatic transition caused by Majornana moving,
which is relevant to
the braiding operations in topological quantum computation.
The scheme is largely based on a time dependent
single-electron-wavefunction approach to quantum transport.
Applying the Kitaev model, we simulate the time dependent Andreev-reflection current
and examine the feasibility of using the current to infer the nonadiabatic transition.
We design a scheme to determine the Landau-Zener tunneling ratio
in the context of transport,
and compare it with the result obtained from the isolated quantum wire.
Desirable agreements are demonstrated for the proposed scheme.
\end{abstract}


\maketitle

\section{Introduction}

The nonlocal nature of the Majorana zero modes (MZMs)
and the intrinsic non-Abelian braiding statistics,
both implying immunity from the influence of local environmental noises,
promise a sound potential for topological quantum computation
\cite{Kit01,Kit03,Sar08,Ter15,Sar15,Opp20}.
The braiding operation is actually exchanging the Majorana modes in real space,
which leads to a unitary rotation in the degenerate subspace of ground states.
The unitary rotation can constitute desired quantum information processing
and realize logic gates in topological quantum computation.

For braiding operations, the early and representative scheme
is quantum-adiabatically moving the MZMs by tuning a series of electric gates
to drive different regions of the Majorana quantum wire
into the topological or non-topological regime \cite{Fish11,Tew11a,Opp12,Plug17},
being guided by the fact that the MZMs will form at the boundaries
between the topological and non-topological regions.
The subsequent alternative proposals include tuning the couplings
between Majorana modes directly or indirectly via modulating
the charging energy on the Majorana island or through quantum dots
\cite{Flen11,Tew11,Bee12,Sato15,Opp15,Fre16,Nay16,Sau16,Sar17,Mora18,Xie21},
measurement-only schemes \cite{Nay08,Los16,Fre17,Fu17}, and others \cite{Naz21}.
We notice that, with the progress of gating control techniques,
the Majorana moving schemes have gained renewed interests
in the past years \cite{Ali18,Roy19,Kel19,MacD20,Byr21}.
In order to realize the topological protection,
i.e., to restrict the quantum evolution
in the subspace of the ground states,
the quantum moving should be adiabatically slow.
However, this may contradict other requirements
such as avoiding the quasiparticle-poisoning decoherence.
In practice, the adiabatic condition may be violated by finite braiding rates,
and the effects of nonadiabatic transition constitute thus an important subject
among the studies \cite{Ali18,Sar11,Opp13,Shn13,Karz14,Fran17,Sau19}.

\begin{figure}
  \centering
  \includegraphics[scale=0.8]{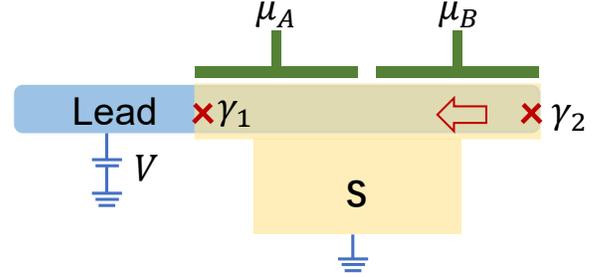}
  \caption{
Schematic setup for transport probe of the nonadiabatic transition caused by Majorana moving,
via measuring the Andreev reflection current flowing back to the lead
from the superconductor through the grounded terminal.
The electrochemical potential $\mu_A$ is tuned to make the left half wire
in topological regime,
while $\mu_B$ is changed to make the right half wire
from initially topological to finally non-topological regimes.
As a consequence, the Majorana mode $\gamma_2$
is moved from the right side to the center of the quantum wire.  }   \label{sys}
\end{figure}

Rather than simulating the nonadiabatic effects
in the isolated quantum wires \cite{Ali18,Sar11,Opp13,Shn13,Karz14,Fran17,Sau19},
we raise the question
how to probe the nonadiabatic transition by experimental measurements.
This interesting problem, to the best of our knowledge,
has not been studied so far in literature.
In this work, we will propose a transport probe scheme to infer
the Majorana-moving-caused nonadiabatic transition,
as schematically shown in Fig.\ 1,
which is actually the minimal setup
of tunneling spectroscopy of Majorana conductances.

Conceptually, the total lead current flowing through the tunnel junction
from the lead to the quantum wire
is the sum of the electron and hole currents,
while the Andreev-reflection current flowing back to the transport lead
from the quantum wire through the grounded terminal
is the double of the hole current.
The Andreev current is proportional to the occupation of the zero-energy quasiparticle,
in this work which will be referred to as
occupation of the Majorana zero-energy bound states (ZEBS).
Therefore, in practice, one can extract information of the ZEBS occupation
from the Andreev current. During Majorana moving, the nonadiabatic transition
will reduce the occupation probability of the ZEBS and thus affect the Andreev current.
However, technically, the complexity is twofold.
On one aspect, the quantum moving and transport probing are highly time dependent.
We are thus required to apply a time dependent transport theory.
In this work, we will apply the single-electron wavefunction (SEWF) approach
\cite{SG91,Li09,Li12b,SG16}, which is a time-dependent
generalization of the stationary $S$ matrix scattering theory.
In particular, this approach was recently extended
to the context of superconductor-induced Andreev reflections \cite{Li20}.
On the other aspect, determination of the Landau-Zener tunneling ratio
is nontrivial in the transport probing process.

The remainder of this work is organized as follows.
We first outline in Sec.\ II
the SEWF approach, in contact with the Kitaev lattice model.
Then, in terms of the time-dependent AR coefficient,
we display in Sec.\ III the Andreev currents in three stages
and discuss several remarkable features.
After carrying out a self-consistency examination, in Sec.\ IV
the Landau-zener transition ratio properly determined from transport probe
is compared with the result calculated in the isolated quantum wire.
Full agreement is demonstrated, which is also followed by support
from the whole moving-period transition and under (small) finite bias voltages.
Finally, we summarize the work in Sec.\ V with discussions.

\section{Single Electron Wavefunction Approach}

For the transport setup schematically shown in Fig.\ 1, the total Hamiltonian
can be split into three parts, $H=H_W+H_{\rm lead}+H^\prime$,
i.e., the sum of the Hamiltonians
of the quantum wire, the transport lead, and their coupling.
The Hamiltonian of the transport lead can be simply described by
$H_{\rm lead}=\sum_l \epsilon_l a^{\dagger}_l a_l$,
with $a^{\dagger}_l$ ($a_l$) the creation (annihilation)
operator of the lead electron with energy $\epsilon_l$.
In this work, we use $l$ to label the continuum of the lead electron states.
The tunnel-coupling Hamiltonian between the quantum wire and the lead is described by
$H'=  \sum_l t_l  c^{\dagger}_1 a_l +\rm{h.c}$,
where $c^{\dagger}_1$ is the electron operator
of the most left site of the quantum wire coupling to the lead.
Without loss of essential physics, we employ in this work the Kitaev model
(spinless $p$-wave superconductor) to describe the quantum wire.
For the $s$-wave superconductor proximitized Rashba nanowire,
one can follow the same treatment developed in this work.
The one dimensional Kitaev lattice model reads as \cite{Kit01}
\bea
H_W &=& \sum_j \left[
-\mu_j c^{\dg}_{j}c_j - \frac{\Omega}{2}  (c^{\dg}_{j}c_{j+1}+{\rm h.c.}) \right] \nl
&& + \frac{\Delta}{2} \sum_j (c_{j}c_{j+1}+{\rm h.c.})  \,.
\eea
$\mu_j$ is the chemical potential which can be tuned via electric gates,
$\Delta$ is the order parameter of superconductor,
and $\Omega$ is the hopping energy between the nearest neighbor sites
with $c^{\dagger}_j$ ($c_j$) the associated electron creation (annihilation) operators.
Introducing the Nambu operator
$\hat{\Psi}=(c_1,\cdots,c_N,c^{\dg}_1,\cdots,c^{\dg}_N)^T$,
the Hamiltonian of the quantum wire can be rewritten as
$H_W=\frac{1}{2}\hat{\Psi}^{\dg}\widetilde{H}_W\hat{\Psi}$,
with $\widetilde{H}_W$ the well-known Bogoliubov de-Gennes (BdG)
Hamiltonian matrix which gives directly the energy spectrum of
the Bogoliubov quasiparticles after diagonalization \cite{Li20}.
Instead of the operators basis,
the BdG Hamiltonian matrix can be understood as well
as constructed under the single-particle state basis
$\{ |e_1\ra,\cdots,|e_N\ra;~ |h_1\ra,\cdots,|h_N\ra \}$,
where $|e_j\ra$ and $|h_j\ra$ describe, respectively,
the electron and hole states on the $j$th site of the quantum wire.
Using the electron and hole basis states,
the tunnel-coupling Hamiltonian can be rewritten as
\bea\label{tunnel-H}
H'=\sum_l t_l (|e_1\ra\la e_l|-|h_1\ra\la h_l|)
   +{\rm h.c.} \,.
\eea
Here, in order to account for the Andreev reflection (AR) process,
we have introduced also the electron and hole states
$\{|e_l\ra, |h_l\ra \}$ for the transport lead.

Following the single particle $S$ matrix scattering approach
within the Landauer-B\"uttiker formalism,
we notice that an electron entering the superconductor
will either excite a Bogoliubov quasipaticle or destroy it,
depending on the quasipaticle state empty or occupied.
The former case corresponds to the normal tunneling process
and the latter results in the Andreev reflection.
After accounting for this physics,
the tunnel-coupling Hamiltonian is truncated as \cite{Li20}
\bea\label{tunnel-H4}
H'=
\sum_l t_l (|\tilde{e}_1\ra\la e_l|-|\tilde{h}_1\ra\la h_l|) + {\rm h.c.} \,,
\eea
where the edge states $|e_1\ra$ and $|h_1\ra$ are projected onto the
subspace of the Bogoliubov quasiparticle states
through $|\tilde{e}_1\ra = \hat{P}|e_1\ra $ and $|\tilde{h}_1\ra = \hat{P}|h_1\ra$.
The projection operator is given by $\hat{P}=\sum^{\prime}_n |E_n\ra\la E_n|$,
with the sum ranging over all the eigenstates of the particle sector
of the BdG Hamiltonian matrix.

In this work, we would like to employ the single-electron wavefunction (SEWF)
approach \cite{SG91,Li09,Li12b,SG16,Li20}
for the highly time-dependent transport problem under study,
which is actually a time-dependent generalization
of the stationary $S$ matrix scattering theory.
The basic idea of the SEWF approach is keeping track of the quantum evolution
of a single electron initially in the lead,
and computing the time-dependent transport coefficients.
Let us assume the electron initially in the lead state
$|\Psi(0)\ra=|e_{\bar{l}}\ra$, and with the incident energy $E_{in}=E_{\bar{l}}$.
The subsequent evolution with time will result in a quantum superposition
of all basis states of the lead and the quantum wire, as
\bea \label{psi}
|\Psi_{\bar{l}}(t)\ra &=& |\Psi_w(t)\ra + |\Psi_{\rm leads}(t)\ra  \nl
&=& \sum^N_{j=1} [ b_{je}(t)|e_j\ra+b_{jh}(t)|h_j\ra ] \nl
&&+ \sum_l [ \beta_l(t)|e_l\ra+\tilde{\beta}_l(t)|h_l\ra ]  \, .
\eea
Substituting it into the time-dependent Schr\"odinger equation
$i\hbar |\dot{\Psi}\ra = H |\Psi\ra$
and applying the technique of Laplace and inverse-Laplace transformations
to eliminate the degrees of freedom of the lead, after some algebras,
we obtain the equations for the quantum wire state as \cite{Li20}
\begin{eqnarray} \label{result-1}
i\hbar \begin{bmatrix}
\dot{b}_{1e} \\ \dot{b}_{2e}\\ \vdots \\\dot{b}_{Ne}  \\
\dot{b}_{1h} \\ \dot{b}_{2h}\\ \vdots \\\dot{b}_{Nh}  \end{bmatrix}
= \left[ \widetilde{H}_W + \left( \hat{P} \Sigma \hat{P}  \right)\right]
\begin{bmatrix}
b_{1e} \\ b_{2e}\\ \vdots \\ b_{Ne}  \\
b_{1h} \\ b_{2h}\\ \vdots \\  b_{Nh}  \end{bmatrix}
+ t_{L} e^{-\frac{i}{\hbar} E_{\rm in}t} \hat{P}
\begin{bmatrix}
1 \\ 0 \\ 0 \\ 0 \\ \vdots \\ 0 \\ 0 \\ 0 \\ 0  \end{bmatrix}
\end{eqnarray}
Here we have introduced the self-energy operator
\bea\label{result-2}
&&\Sigma = (-i\Gamma_L/2)\,(|e_1\ra\la e_1|+|h_1\ra\la h_1|) \,.
\eea
The tunnel-coupling rate under the wide-band limit reads as $\Gamma_L=2\pi\rho_L t_L^2$,
with $\rho_L$ the density-of-states of the transport lead,
and $t_L$ the constant tunnel-coupling strength (i.e., $t_l= t_L$).

The single ($\bar{l}$) electron currents are obtained through the changing rates
of the occupation probabilities of the lead electron and hole, given by
\bea\label{result-4}
i_{\rm Le}&=& -\frac{e}{\hbar}\frac{\partial P_{\rm Le}}{\partial t}
= - \frac{e}{\hbar}\frac{\partial \sum_l|\beta_{l}(t)|^2}{\partial t}  \,,  \nl
i_{\rm Lh}&=& \frac{e}{\hbar}\frac{\partial P_{\rm Lh}}{\partial t}
=\frac{e}{\hbar}\frac{\partial \sum_l|\tilde{\beta}_{l}(t)|^2}{\partial t}  \,.
\eea
One may note that $i_{\rm Le}$ and $i_{\rm Lh}$ are, respectively,
the electron and hole currents
flowing through the tunnel junction from the lead to the quantum wire.
Moreover, after some algebras, one can further obtain
\bea\label{result-3}
i_{\rm Le} &=& -\frac{e}{\hbar}
\left[ 2t_L {\rm Im}(e^{\frac{i}{\hbar}E_{in}t}\la e_1|\hat{P}|\Psi_w\ra)
+\Gamma_L\, |\la e_1|\hat{P}|\Psi_w\ra|^2  \right]     \nl
i_{\rm Lh} &=& \frac{e}{\hbar}\Gamma_L\, |\la h_1|\hat{P}|\Psi_w\ra|^2  \,.
\eea
We are thus allowed for very convenient computation for the time-dependent currents
by solving \Eq{result-1}
for the time-dependent state $|\Psi_w(t)\ra$ of the quantum wire.
In steady state, $i_{\rm Le}=i_{\rm Lh}$.

Along the same line of the Landauer-B\"uttiker formalism
of single-particle scattering theory,
the currents under finite bias voltage (at zero temperature) can be computed as
\bea\label{current-1}
I_{\rm Le} &=& \int_{-\mu_L}^{\mu_L} d E_{\bar{l}}\rho_L i_{\rm Le}
= \frac{e}{h}\int_{-\mu_L}^{\mu_L} d E_{\bar{l}} {\cal T}_{L} \,,  \nl
I_{\rm Lh} &=& \int_{-\mu_L}^{\mu_L} d E_{\bar{l}}\rho_L i_{\rm Lh}
= \frac{e}{h}\int_{-\mu_L}^{\mu_L} d E_{\bar{l}} {\cal T}_{A}  \,.
\eea
Here, we have related the results from the single-particle currents
with the conventional theory in terms of transmission coefficients.
However, the present SEWF approach is a time-dependent generalization of
the Landauer-B\"uttiker $S$ matrix scattering theory
and also of the nonequilibrium Green's function treatment.
Note also that, the total lead current
flowing through the tunnel junction from the lead to the quantum wire
is the sum of the electron and hole currents,
i.e, $I_{\rm L}=I_{\rm Le}+I_{\rm Lh}$,
while the Andreev current flowing back to the transport lead
from the quantum wire through the grounded terminal
is $I_{\rm A}=2I_{\rm Lh}$.
Accordingly, the linear Andreev differential conductance
is given by $G_{A}=(2e^2/h){\cal T}_{A}$.
The Andreev reflection coefficient ${\cal T}_{A}$
can be calculated through
\bea\label{i-T}
{\cal T}_{A}=\frac{h}{e}\rho_L i_{\rm Lh}
=\frac{\Gamma^2_L}{t^2_L}\, |\la h_1|\hat{P}|\Psi_w\ra|^2 \,.
\eea
One can prove that, in steady state, this result will recover the standard expression
in terms of the nonequilibrium Green's functions \cite{Li20}.

\section{Time Dependent Current and Self-Consistency Examination}

The time-dependent SEWF approach outlined above is in particular suitable
for simulating the transport probe of nonadiabatic transitions
\cite{Ali18,Sar11,Opp13,Shn13,Karz14,Fran17,Sau19}.
In this work, we consider the simplest
piano-key-model analyzed in Ref.\ \cite{Ali18}.
Let us denote the electrochemical potentials of the left
and right half wire by $\mu_A$ and $\mu_B$.
By fixing $\mu_A$ with a value smaller than $\Omega$,
the left half is kept in the topological regime,
and the right half is tuned according to \cite{Ali18}
\begin{equation}\label{mu-b}
\mu_B(t)=[1-f(t/\tau)]\mu_{Bi}+f(t/\tau)\mu_{Bf}  \,,
\end{equation}
where the initial value of $\mu_B$
is set as $\mu_{Bi}=\mu_A$
and the final value $\mu_{Bf}$ is larger than $\Omega$.
In general, $f(s)$ can be a monotonically increasing function
with $f(0) = 0$ and $f(1) = 1$.
As an example, one can choose $f(s) = \sin^2(s\pi/2)$.
Therefore,
the right half of the quantum wire begins in a topological regime
with $\mu_{Bi}=\mu_{A}<\Omega$,
and finishes at time $\tau$ with $\mu_{Bf}>\Omega$,
which drives the right half wire into a trivial regime.
As the consequence, the Majorana bound state
at the right end of the wire will be moved to the center.

\begin{figure}
  \includegraphics[scale=0.4]{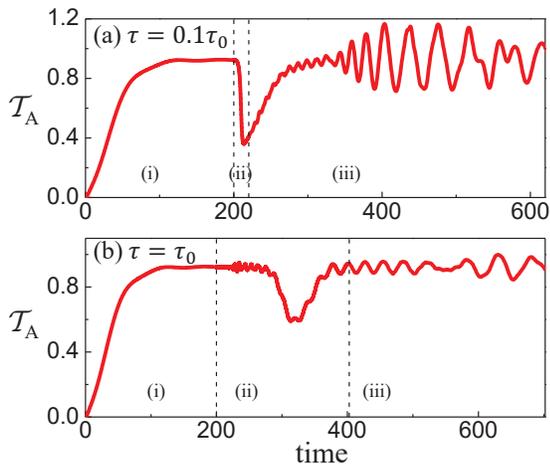}
  \caption{
Time-dependent behaviors of the Andreev reflection current
(in terms of the time-dependent AR coefficient ${\cal T}_A$)
associated with three stages:
(i) steady-state formation of the ZEBS occupation
before switching on the gate voltage ($\mu_B$) control for Majorana moving;
(ii) gate-control Majorana moving; and (iii) after the moving.
In the simulation, we set the nearest-neighbor hopping energy $\Omega=1$
as the energy unit in an arbitrary system of units.
Accordingly, other parameters are set as:
$\Delta=0.5$, $\mu_A=0.9$, $\mu_B=(0.9, 1.1)$,
$\Gamma_L=1.0$, and $E_{in}=0.0$.
The moving times are chosen as $\tau=0.1\tau_0$ in (a) and $\tau=\tau_0$ in (b),
while the Landau-Zener characteristic time $\tau_0$ is determined by \Eq{tau-0}
in the main text, which gives $\tau_0=202.64$.   }\label{Fig1}
\end{figure}

As a preliminary illustration, we display in Fig.\ 2 the time-dependent behaviors
of the Andreev current (in terms of the time-dependent AR coefficient ${\cal T}_A$)
associated with three stages:
(i) steady-state formation of the ZEBS occupation
before switching on the gate ($\mu_B$) control for Majorana moving;
(ii) gate-control Majorana moving; and (iii) after the moving.
In Fig.\ 2,
we first illustrate the gradual formation of steady-state current during stage (i),
which indicates the formation of stationary ZEBS occupation.
The formation of steady-state current
starting with a single $\bar{l}$-lead-electron is somehow a bit tricky.
In a simple way, we can say that
it corresponds to the long time limit of the $S$ matrix scattering approach,
As a more detailed inspection, in \Eq{result-1}
the initial condition with the $\bar{l}$ electron injection
manifests as a driving term, which will result in
nonzero occupation of the wire states
and nonvanishing continuous current, even in long time limit.
This interesting feature differs from the alternative initial condition
by considering the electron initially in the quantum wire,
e.g., on the zero-energy bound state,
which will unavoidably lose the electron into the continuum of the lead reservoir.
This issue can be understood better in the simpler transport problem
by considering an electron transmitting through a resonant level of quantum dot \cite{SG16}.
After a careful check, one can find that
the single $\bar{l}$ electron current is extremely small,
by noting that the coupling (energy) amplitude ($t_l$ in \Eq{tunnel-H})
is infinitesimally small,
because of the infinite extension of the lead electron wavefunction
in the continuum of eigenstates representation.
Therefore, only after multiplying the density-of-states of the lead,
which is infinity for a continuum,
the finite result of current
or the transmission coefficient as shown in \Eq{i-T} can be obtained.
We can thus conclude that
the steady-state single $\bar{l}$ electron current,
which is infinitesimally small and flows for infinitely long time,
does not contradict any physical principle.
After multiplying the density-of-states of the lead
and integrating over some energy range,
the single electron transmission approach matches well the
transport problem under bias voltage, where the external
circuit will help to maintain the continuous transport current.

During the stage (ii) as shown in Fig.\ 2,
modulating $\mu_B$ from $\mu_{Bi}$ to $\mu_{Bf}$ according to \Eq{mu-b}
moves the Majorana bound state $\gamma_2$
from the right end to the center of the quantum wire.
Two moving times, $\tau=0.1\tau_0$ and $\tau_0$, are chosen for the simulation.
Here, the characteristic time $\tau_0$ is determined
from the Landau-Zener tunneling analysis \cite{Ali18}, which reads
\begin{equation}\label{tau-0}
\tau_0=|\mu_{Bf}-\mu_{Bi}|\left(\frac{N_{right}}{\pi \Delta }\right)^2 \,,
\end{equation}
with $N_{right}$ the lattice number of the right half wire.
Notice that, during the continuous modulation of $\mu_B$,
the superconductor energy gap
will almost close at $\mu_B=\Omega$ and reopen after $\mu_B>\Omega$.
The closing and reopening of the energy gap
is associated with a topological phase transition.
Accordingly, the nonadiabatic transition is expected to take place mainly
when crossing the phase transition point.
In Fig.\ 2(a), for the fast moving with $\tau=0.1\tau_0$,
after crossing the phase transition point,
the short time does not allow obvious re-occupation of the ZEBS
from injection of the lead electron,
resulting thus in the decreasing Andreev current as observed.
In contrast, In Fig.\ 2(b), the slower moving (with $\tau=\tau_0$)
reveals a re-occupation behavior
after crossing the phase transition point,
as indicated by the turnover behavior of the current.

In Fig.\ 2, we also show the current behavior
in stage (iii) after the moving.
In this context, we notice two interesting features.
One is the current oscillation,
which becomes stronger after the fast moving;
another is that the AR coefficient ${\cal T}_A$ can be larger than one.
The reason for the former is that
the occupation of the excited quasiparticle states
caused by the nonadiabatic transition
will interfere with the re-occupation of the ZEBS owing to new injection,
since the AR current is extracted from the left-end-site of the quantum wire,
which is commonly shared by the ZEBS and the exited states.
The occupation of the excited quasiparticle states
is also the reason for the second phenomenon mentioned above.
This corresponds to an increase of transport channels.
From the Landauer-B\"uttiker scattering theory (the multi-channel version),
we know that the transport coefficient can be larger than one.

\begin{figure}
  \includegraphics[scale=0.35]{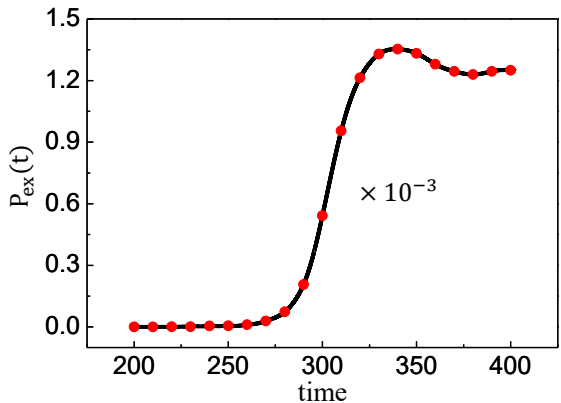}
  \caption{
Self-consistency examination for using the net accumulation of currents
to infer the change of occupation probabilities of the wire states.
Illustrated in this plot is the total probability
of transition to the excited states.
Here we consider the contribution of multiple electrons
near the incident energy $E_{in}$,
within a small energy window $\Delta E=\alpha \widetilde{\Gamma}_L$
by setting $\alpha=10^{-2}$,
while $\widetilde{\Gamma}_L$ is the tunnel-coupling rate
(widening width) of the ZEBS.
$\widetilde{\Gamma}_L$ can be determined as $\widetilde{\Gamma}_L=|u_1|^2\Gamma_L$,
with $u_1$ the electron amplitude of the first lattice site (coupled to the lead)
in the wavefunction of the ZEBS, which can be obtained
by diagonalizing the BdG Hamiltonian of the quantum wire.
All the parameters used here are the same as in Fig.\ 2.    }
\end{figure}

In the SEWF approach, the net accumulation
of the normal tunneling current $i_{\rm Le}$
and the Andreev process current $i_{\rm Lh}$
should correspond to
the change of the occupation probability of the ZEBS
plus the transition to other excited states. Thus we have
\begin{equation}\label{ie-ia}
\frac{1}{e}\int_{t_i}^{t} \left[i_{\rm Le}(t')-i_{\rm Lh}(t')\right]\,dt'
= [P_{\rm E_0}(t)-P_i] + P_{\rm ex}(t) \,.
\end{equation}
In the numerical simulation, we can compute $P_{\rm E_0}(t)$ as
$P_{\rm E_0}(t)=|\la E_0|\Psi_w(t)\ra |^2$, and $P_{\rm ex}(t)$ as
\bea\label{Pex}
P_{\rm ex}(t)&=& \sum_j (|\la E_j(t)|\Psi_w(t)\ra |^2+|\la -E_{j}(t)|\Psi_w(t)\ra |^2)  \nl
&& - [P_{\rm E_0}(t)-P_i]  \,,
\eea
where the summation is
over all the instant eigenstates of the BdG Hamiltonian matrix.
As an examination for self-consistency, we compare in Fig.\ 3
the probability $P_{\rm ex}(t)$ computed through \Eq{Pex}
and the result from \Eq{ie-ia}.
Notice that, as explained above, the currents and the occupation probabilities
from a single $\bar{l}$ electron on the two sides of \Eq{ie-ia}
are infinitesimally small.
In the plot of Fig.\ 3, we consider the contribution of multiple electrons
near the energy $E_{\bar{l}}$ of the $\bar{l}$ electron,
within a small energy window $\Delta E=\alpha \widetilde{\Gamma}_L$
by setting $\alpha=10^{-2}$,
while $\widetilde{\Gamma}_L$ is the tunnel-coupling rate
(widening width) of the ZEBS.
Note also that in the numerical simulation we set $\Gamma_L=2\pi\rho_L |t_L|^2=1$.
The full agreement in Fig.\ 3 indicates that
we can employ the currents and their accumulations
to infer the Majorana-moving caused nonadiabatic transition,
as to be analyzed below in detail.

\section{Landau-Zener Ratio Inferred By Transport Probe}

We may regard the Majorana-moving-caused nonadiabatic transition
as the Landau-Zener tunneling.
For an isolated quantum wire, if knowing the initial and final
occupation probabilities of the ZEBS, $P_i$ and $P_f$,
we can unambiguously define the Landau-Zener transition ratio
as $\gamma=\Delta P/P_i$, where $\Delta P=P_i-P_f$,
since this loss of the occupation probability
is owing to the nonadiabatic transition.
However, for the transport probing, the problem is more complicated.
After the first stage, i.e., stage (i) as shown in Fig.\ 2,
steady-state occupation of the ZEBS is achieved.
Then, in stage (ii), during the time period $(t_i,t)$
(with $t<t_f$ and $\tau=t_f-t_i$),
net accumulation of the quasiparticle-states occupation is the difference
of the normal electron injection and the AR loss, while the both are given by
\bea
a &=& \frac{1}{e}\int^{t}_{t_i} dt'\, i_{\rm Le}(t')  \,,  \nl
b &=& \frac{1}{e}\int^{t}_{t_i} dt'\, i_{\rm Lh}(t')  \,.
\eea
We may thus introduce $\widetilde{P}_i=P_i +(a-b)$
and use it to replace $P_i$,
for the ``initial" occupation probability on the ZEBS.
However, for moving after crossing the phase transition point,
the net accumulation is largely not participating in the nonadiabatic transition,
owing to the reopening energy gap during the later half time period $(t_i+\tau/2, t_f)$.
This insight suggests us to introduce
\bea
\tilde{a} &=& \frac{1}{e}\int^{t_i+\tau/2}_{t_i} dt'\, i_{\rm Le}(t')  \,,  \nl
\tilde{b} &=& \frac{1}{e}\int^{t_i+\tau/2}_{t_i} dt'\, i_{\rm Lh}(t')  \,,
\eea
to replace the quantities $a$ and $b$ in $\widetilde{P}_i$
after crossing the phase transition point.
That is, when $t>t_i+\tau/2$, we use $\widetilde{P}_i = P_i +(\tilde{a}-\tilde{b})$.
For all, we define the Landau-Zener transition ratio $\widetilde{\gamma}$ as
\bea
\widetilde{\gamma} = \frac{\Delta\widetilde{P}}{\widetilde{P}_i}  \,,
\eea
where the transition probability is given by
$\Delta\widetilde{P}=(P_i-P_f)+(a-b)$.
Note that here the ``final" occupation probability $P_f$
is defined at the running time $t$.

\begin{figure}
  \includegraphics[scale=0.65]{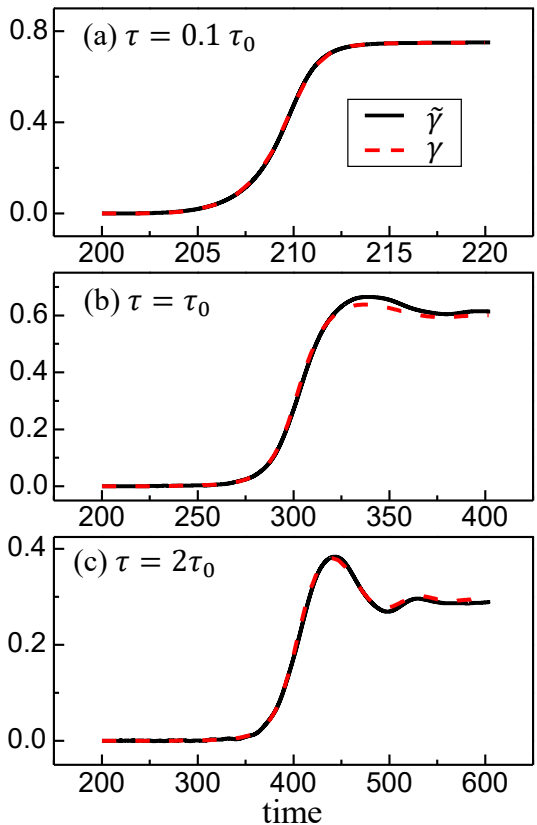}
  \caption{
Landau-Zener transition ratio during the moving process,
in comparison between $\widetilde{\gamma}$ (inferred by transport probe)
and $\gamma$ (calculated for the isolated quantum wire).
Moving times of $\tau=0.1\tau_0$, $\tau_0$ and $2\tau_0$
are considered, respectively, in (a), (b) and (c).
Other parameters are the same as in Fig.\ 2.   }
\end{figure}

In Fig.\ 4, we display the numerical results of $\widetilde{\gamma}$ versus $\gamma$
for different moving speeds and find desirable agreements.
In the numerical simulation, we determine $P_i$ and $P_f$
from the wavefunction of the quantum wire by solving \Eq{result-1}.
In practice, they should be extracted from the Andreev current
$i_{\rm A}(t) \simeq (2e)\,\widetilde{\Gamma}_L P_{\rm E_0}(t)$,
where $\widetilde{\Gamma}_L$ is the coupling rate of the ZEBS to the lead
and the current is recorded
by applying a small bias voltage (in the linear response regime).
In this extracting protocol, $P_i$ can be determined rather precisely,
since there is no nonadiabatic transition to the excited states
until the formation of steady state.
$P_f(t)$ is expected also to be determined
with reasonable precision from $i_{\rm A}(t)$,
provided that $\widetilde{\Gamma}_L$ is much smaller than the energy gap
and the quantum moving is not so fast.

We have shown in Fig.\ 4
the transient transition behavior during the moving process.
Now we further show in Fig.\ 5 the final result of transition
after the entire $\tau$-period quantum moving, which corresponds to
the standard characterization of the Landau-Zener tunneling.
It is well known that the Landau-Zener tunneling formula
predicts exponential dependence on the moving time $\tau$.
We see here that,
in addition to the good agreement between $\widetilde{\gamma}$ and $\gamma$,
the exponential dependence behavior maintains well for wide range of $\tau$,
despite that the present problem is beyond the simple
two-level system considered when deriving the Landau-Zener formula.
We notice certain deviation only for fast moving (small $\tau$),
which is reasonable by noting that, in the limit of $\tau\to 0$,
the ``transition" is actually from a transformation of the basis states
that does not obey the Landau-Zener transition formula.

\begin{figure}
  \includegraphics[scale=0.43]{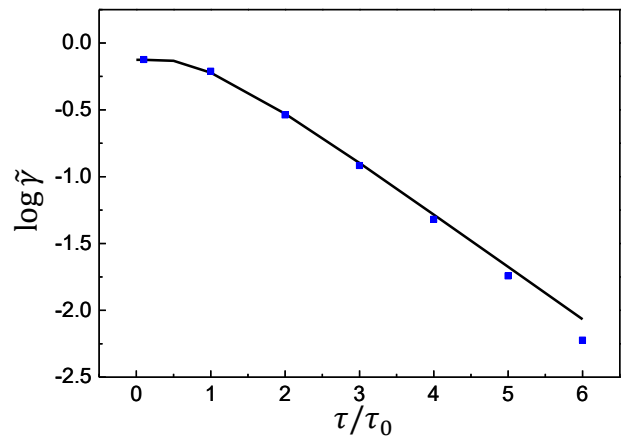}
  \caption{
Landau-Zener transition ratio as a function of the moving time ($\tau$).
The results plotted by the filled squares
are the ratio $\widetilde{\gamma}$ inferred by transport probe,
which are compared with the results (solid-line) calculated for the isolated quantum wire.
Parameters are the same as in Fig.\ 2.   }    \label{Fig3}
\end{figure}

\begin{figure}[h]
  \includegraphics[scale=0.4]{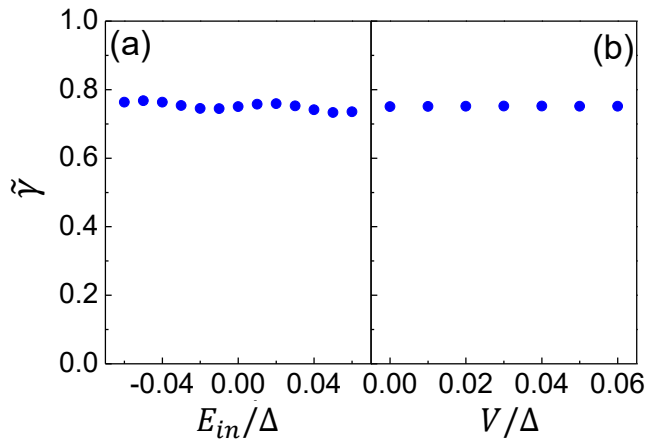}
  \caption{
(a) Landau-Zener transition ratio probed
by electrons with different incident energies.
(b) Averaged transition ratio for
transport probe under (small) finite bias voltages.
In this plot we consider the moving time $\tau=0.1\tau_0$.
Other parameters are the same as in Fig.\ 2.  }    \label{Fig4}
\end{figure}

As explained previously, the single electron approach can account well for
the contribution of multiple electrons at the incident energy $E_{in}=E_{\bar{l}}$,
by multiplying a proper constant (the lead electron numbers at this energy).
We can thus apply the SEWF approach to compute the differential conductance,
which is directly related to the transmission coefficient at zero temperature,
and/or compute the linear response current.
In practice, one may apply some finite (but also small) bias voltage.
In this case, we expect the transport-probe protocol
discussed in this work to be valid as well.
Indeed, in Fig.\ 6(a), we show the transition results probed
by electrons with different incident energies.
And, in Fig.\ 6(b), we average the results to obtain the transition ratio,
which corresponds to the transport probe under different bias voltages.
The result is found insensitive to the (small) bias voltages.
This observation can benefit the feasibility of the proposed transport probing scheme.

\section{Summary and Discussions}

We proposed a scheme to probe the Majorana-moving-caused nonadiabatic effects.
Our analysis was based on a recently developed single-electron-wavefunction approach,
which generalizes the stationary BdG $S$ matrix scattering theory
to a time-dependent transport version, being thus in particular
suitable for the proposed highly time-dependent transport problem.
Via numerical simulation with the Kitaev model, we displayed and discussed
the time-dependent behaviors of the Andreev currents
flowing back to the lead through the grounded terminal of the superconductor,
during the process of Majorana moving.
We further examined the feasibility of using the current
to infer the nonadiabatic transition
and designed a scheme to determine
the Landau-Zener tunneling ratio in the context of transport.
By comparing with the results of isolated quantum wire,
we demonstrated desirable agreements
for both the transient and final Landau-Zener tunneling ratios,
together also with simulations under (small) finite bias voltages.

Majorana moving and the nonadiabatic effects are crucial ingredients of braiding dynamics.
Understanding the moving effects on Majorana bound states and quasi-particle excitations
is a prerequisite for Majorana manipulation with high fidelity,
since the manipulation at finite timescales will cause
both decoherence and renormalization effects.
Important extension of the present study is to consider the typical realization of
the Rashba semiconductor nanowire in proximity to an $s$-wave superconductor,
while the Kitaev model simulated in this work
can be regarded as the strong magnetic field limit.
For such more complicated quantum wires,
the Landau-Zener characteristic time $\tau_0$
may exist significant uncertainty compared with estimate values
and may have $1\sim 2$ orders of magnitude change with the system parameters \cite{Ali18}.
Moreover, in realistic quantum wires, Majorana moving
passing through static impurities/disorders present in the system
may have essential influence on the nonadiabatic transitions,
as qualitatively discussed in Refs.\  \cite{Opp13,Shn13}.
Including such complexities in our time dependent
lattice-model based scheme of simulations should be straightforward.
However, further inclusion of
inelastic scattering effects, quasi-particle excitation and poisoning
is important but more challenging for theoretical simulations.
In this work, we only considered
the simplest moving scheme analyzed in Ref.\ \cite{Ali18}.
It will be valuable to simulate the transport probe
of nonadiabatic effects associated with other moving schemes,
e.g., the domain-wall moving \cite{Opp13,Shn13}
and the optimal control of quantum moving
to suppress the nonadiabatic transition \cite{Karz14}.
Important issues associated with large moving speed
may include the stability of the Majorana bound states
and possible relativistic effects \cite{Opp13,Shn13}.
Simulating all these effects in connection with transport probe
is of great interest for future studies.

\vspace{1.5cm}
{\flushleft\it Acknowledgements.}---
This work was supported by the
National Key Research and Development Program of China
(No.\ 2017YFA0303304) and the NNSF of China (Nos.\ 11675016 \& 11974011\&11904261).


\end{CJK*}
\end{document}